\documentclass[review,number,sort&compress]{elsarticle}
\usepackage{hyperref}
\usepackage{breakurl}
\usepackage{graphicx}
\usepackage{amsmath}
\usepackage[english]{babel}

\journal{NIM A}
\begin{document}
\begin{frontmatter}
\title{Application of a Portable $^3$He-Based Polarization Insert at a Time-of-Flight Neutron Reflectometer}

\author[a]{Wolfgang Kreuzpaintner}
\ead{Wolfgang.Kreuzpaintner@frm2.tum.de}

\author[b]{Sergey Masalovich}
\author[c]{Jean-Fran\c{c}ois Moulin}
\author[a]{Birgit Wiedemann}
\author[a]{Jingfan Ye}
\author[a]{Sina Mayr}
\author[a]{Amitesh Paul}
\author[c]{Martin Haese}
\author[c]{Matthias Pomm}
\author[a]{Peter B\"oni}

\address[a]{Technische Universit\"at M\"unchen, Physik-Department E21, James-Franck-Str.\ 1, 85748 Garching, Germany}
\address[b]{Technische Universit\"at M\"unchen, Forschungsneutronenquelle Heinz Maier-Leibnitz \big(MLZ, FRM\,II\big), Lichtenbergstr.\ 1, 85747 Garching, Germany}
\address[c]{Helmholtz-Zentrum Geesthacht, Zentrum f\"ur Material und K\"ustenforschung, Au\ss enstelle am MLZ in Garching bei M\"unchen, Lichtenbergstr.\ 1, 85748 Garching, Germany}

\begin{abstract}
The suitability of a transportable $^3$He-spin filter as temporary broadband polarizer for a Time-of-Flight neutron reflectometer is demonstrated. A simple two-wavelength method for characterisation of a $^3$He-spin filter is proposed, which can be applied even if the absolute transmittance of the $^3$He-spin filter cannot be accurately determined. We demonstrate the data treatment procedure for extracting the spin-up and spin-down neutron reflectivity from measurements obtained with a time dependent $^3$He polarization. The extraction of a very weak magnetic signal from reflectivity data, measured on the \textit{in-situ} grown  magnetic heterostructure Fe$_{\text{1\,nm}}$/Cu$_{\text{20\,nm}}$/Si$_{\text{substrate}}$ in an externally applied magnetic field of 30\,mT is presented and compared to similar measurements on the growth stage Cu$_{\text{20\,nm}}$/Si$_{\text{substrate}}$ of the very same sample, which does not yet contain any magnetic material.
\end{abstract}


\end{frontmatter}


\section{Introduction}

Many neutron reflectometers are particularly laid out for the analysis of biological, polymer or deuterated samples or for the analysis of solid/liquid interfaces \big(see e.\,g.\ \cite{Campbell2011107,Strobl2011055101,Webster20061164}\big). Their design, therefore, does not always include a polarizer as a standard component. However, Polarized Neutron Reflectometry \big(PNR\big) is a constantly growing area in neutron scattering \cite{Cubitt20021,Lauter20092543,Charlton201115,Saerbeck2012081301,Devishvili2013025112,Vorobiev201525,Moulin20151,Mattauch20151,Stahn201644} and the need to at least temporarily implement a polarization option is desirable if only a particular unpolarized reflectometer allows for the installation of a certain sample environment.
In this case, the temporary character of the polarization solution, a fast and zero-alignment installation and an experimental simplicity are the most decisive factors. 
Today, neutron reflectometers often operate in Time-of-Flight \big(ToF\big) mode, where the momentum transfer $q_z=\frac{4\pi}{\lambda}\sin\alpha_i$ is measured at a fixed angle of incidence $\alpha_i$ and a continuous incident neutron beam wavelength \big($\lambda$\big) spectrum from 2\,\AA\ to 20\,\AA\ \big(either as full range or as a part thereof\big). Therefore, the polarizer also has to operate in a broad wavelength band.

Polarized $^3$He neutron spin filters \big(NSFs\big) \big(see e.\,g.\ \cite{Keith1998236, Jones2000772, Chen2003196, Hutanu2012012012, Masalovich2007791,Andersen20061134,Schulz2010012068}\big) have conquered a very important realm in the last two decades.
Their polarizing effect is due to spin-polarized $^3$He having a very large spin-dependent absorption cross-section. Importantly, the NSFs can be optimised for any wavelength range of cold or thermal neutron beam. In combination with a solid state beam polarizer, NSFs are typically used as analyser after the sample position, as they are simple transmittance devices, which do not affect the phase space of the neutron beam and cover a large angular regime. Obviously, a NSF can also be used as the primary beam polarizer. Although it cannot compete with the performance of a solid state polarizer \cite{Georgii2003250,Keller2000474,Saerbeck2012081301}, the application of NSFs looks to be very attractive as they fulfill the already mentioned requirements of fast and zero-alignment installation and easy characterisation, which is not always the case when using a solid state polarizer \cite{Bigault012017,Tong201477}. 

NSFs only became widely available with the progress achieved in the large-scale production of hyperpolarized $^3$He gas, either by Spin Exchange Optical Pumping \big(SEOP\big) \cite{Bouchiat1960373,Chen2007013416} or by Metastability Exchange Optical Pumping \big(MEOP\big) \cite{Colegrove19632561,Batz2005293}. A SEOP polarization setup is typically integrated into a neutron instrument for polarizing a $^3$He gas of high pressure directly in a neutron beam. This provides a constant degree of $^3$He polarization at a single instrument, but with large space requirements and typically high operational costs. In contrast, MEOP is generally used to polarize $^3$He gas in a polarization station away from the neutron instrument. Here, the aim is not to provide a constant polarization for a single instrument, but to supply a large number of neutron instruments with transportable NSFs in parallel. In this aspect, transportable NSFs clearly have their advantages in a fast, easy and nearly zero-effort installation. Additionally, they can be fully characterised by their transmittance $T(\lambda)$ in the course of the actual experiment. 

The drawback of using transportable NSFs as temporary polarizer is their relatively high absorption of neutrons, which scales with the achievable polarization of the $^3$He gas, and the necessity to work with a variable $^3$He gas polarization with decay \big(relaxation\big) times $T_1$ of typically 150 -- 250\,h \big(see e.\,g.\ \cite{Boag2014012019,Deninger2006439}\big). This decay means that the replacement of a NSF is required when performing long running measurements. It must also be taken into account that $T_1$ can be influenced by an external event, such as applying or changing magnetic fields to the sample. An adequate planning of the experimental details and of the measurement sequence is, hence, required. Due to neutron flux limitations on the majority of nowadays neutron instruments, the data acquisition times required for sample measurements cannot be considered as being very short when compared with the decay time of the $^3$He polarization, regardless of the initial $^3$He gas polarization. The gas polarization, therefore, must be monitored and taken into account in the data evaluation.
This paper demonstrates the use of a single transportable NSF \big(which was prepared by the MEOP method\big) in a magneto-static cavity as temporary broadband polarizer for the horizontal ToF neutron reflectometer REFSANS \cite{Kampmann20061161, Kampmann2004E763, Moulin20151} at the Forschungs-Neutronenquelle Heinz Maier-Leibnitz \big(MLZ\big). Both, a simple and universal characterisation method for NSFs and the procedure for extracting the spin-up \big($\uparrow$\big) and spin-down \big($\downarrow$\big) neutron reflectivity $R^{\uparrow}(q_z)$ and $R^{\downarrow}(q_z)$ from the measured neutron intensities, obtained at variable $^3$He polarization over longer data acquisition periods, will be demonstrated step by step. The distinguishing feature of our approach is that only one spin-flip of the $^3$He gas polarization is sufficient to extract the reflectivity data.

Repeated-flipping measurement sequences with a typically high number of flipping processes are often favoured as a simple way to suppress instabilities in an experimental apparatus. This approach is of benefit if external influences (e.\,g.\ from neighbouring beamlines), which might contribute to a faster or even abrupt depolarisation of the $^3$He gas, cannot be ruled out. As every flipping of $^3$He spins with a radio-frequency signal slightly depolarizes the NSF, repeated-flipping requires a rather perfect spin-flipper with little losses in $^3$He polarization on a flipping event. Although a spin-flip efficiency can be as high as 99.99\,\%, spin-flippers with lower efficiency have to be kept in mind for which the number of $^3$He polarization flipping processes should be kept at a minimum. This, however, then requires the full elimination of external influences. Both techniques (multiple or single spin-flip) can be applied for measuring reflectivity curves and a researcher might have to decide for the ideal measurement sequence for a given case. 

We demonstrate a measurement sequence at which the number of $^3$He polarization flipping processes was kept at a minimum. In this context, we also show a data extraction process which is different to the usually applied polarization efficiency corrections \cite{Wildes1999421,Krycka20092561}, as it does not correct different spin-up and spin-down intensities by the spin-leakage, but uses a direct way to extract the sample parameters $R^{\uparrow}(q_z)$ and $R^{\downarrow}(q_z)$, from the measured intensities using the polarized beam of imperfect and time dependent polarization. In this context, even a rather moderate polarization performance outside the optimal NSF's wavelength band of operation can be accepted and only slightly affects the accuracy for $R^{\uparrow}(q_z)$ and $R^{\downarrow}(q_z)$.


The sample for demonstrating the extraction of $R^{\uparrow}(q_z)$ and $R^{\downarrow}(q_z)$ reflectivity curves from the measurements was directly grown in the neutron beam by metal-metal-epitaxy on silicon \cite{Chang199198}, using a prototype of an \textit{in-situ} Ultra-High-Vacuum thin film deposition setup, which will be presented elsewhere. The size and space requirements of this sample environment also exemplarily justify the use of a by default unpolarized reflectometer at a neutron source nearby to demonstrate the temporary polarization technique. By this method, two different growth stages, Cu$_{\text{20\,nm}}$/Si$_{\text{substrate}}$ \big(growth stage 1\big) and Fe$_{\text{1\,nm}}$/Cu$_{\text{20\,nm}}$/Si$_{\text{substrate}}$ \big(growth stage 2\big), of the very same epitaxially grown thin film sample could directly be compared with each other. Due to the different neutron cross-sections for $\uparrow$ and $\downarrow$ neutrons of the magnetic Fe layer in growth stage 2, an overlaid splitting in the reflectivity curve for $\uparrow$ and $\downarrow$ polarized neutrons is expected. Due to the small amount of magnetic material present, the splitting is also expected to be very small. The ability to measure that splitting demonstrates the feasibility of the measuring method and of the data evaluation procedure. Limitations in the applied sample environment, however, rendered the accurate determination of the absolute NSF transmittance impossible. Therefore the well established NSF characterisation procedure, which is based on the knowledge of the absolute transmittance was fully re-thought and modified to purely rely on the relative transmittances at different wavelengths, which are simultaneously measured in ToF mode. It should also be noted that our method can equally well be applied to multiple flipping sequences as an alternative to a linear approximation approach.

\section{Experimental Procedure}
\begin{figure}
	\begin{center}
		\includegraphics[width=0.75\textwidth]{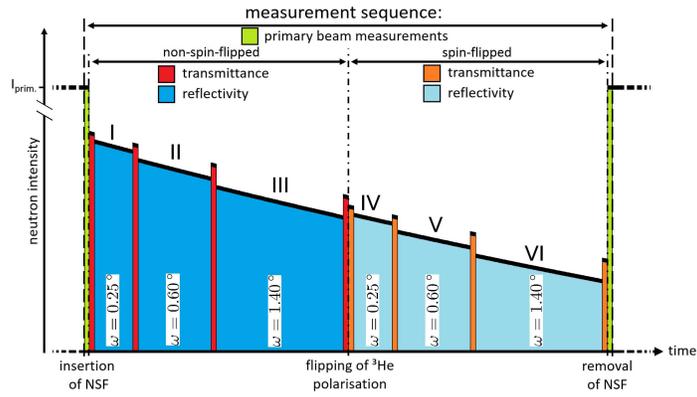}
			\caption{Schematic view of the time dependence of the neutron intensities reaching the detector \big(black line\big), overlaid with the corresponding measurement sequence of NSF transmittance \big(red, orange\big) and sample reflectivity measurements I -- VI \big(blue, light blue\big), flanked by primary beam measurements \big(green\big). With increasing nominal angle of incidence $\omega$ the reflectivity of the sample decreases. In order to rule out any $^3$He depolarization effects caused by spin-flipping, all required non-spin-flipped reflectivities were measured first, followed by the corresponding spin-flipped measurements after a single intermediate flipping of the $^3$He polarization. The duration between the measurements was kept as short as possible and was only defined by the times required for driving the goniometer axes and for restarting the data acquisition \big(typically a few minutes\big). During the long sample reflectivity measurements, the transmittance of the NSF and its polarization efficiency decrease, which must be corrected for in the data evaluation process.}
			\label{Fg:Kreuzpaintner_1}
	\end{center}
\end{figure}

PNR measurements were carried out using a 2\,cm wide vertically collimated neutron beam with the broad wavelength-spectrum of $2.25$\,\AA\ $\leq \lambda \leq 7.75$\,\AA\ introduced in section \ref{subsec:Step 1}. Although a larger wavelength band is possible, the chosen settings cover the optimal intensity yield of the beamline, which the NSFs were also optimised for. The measurements were performed with a wavelength resolution of $3.1$\,\% and at three different nominal angles of incidence \big($\omega=0.25$\,$^{\circ}$, $\omega=0.6$\,$^{\circ}$ and $\omega=1.4$\,$^{\circ}$\big), such that a reflectivity curve up to $q_z \approx 0.1$\,\AA$^{-1}$ was recorded in three overlapping parts. Due to ballistic effects, the nominal angle of incidence $\omega$ slightly differs from the wavelength dependent effective angle of incidence $\alpha_i = \alpha_i(\lambda)$ on the sample and was corrected for in the data analysis process. At each nominal angle of incidence, the height of the neutron beam was adjusted such that the sample was optimally illuminated, whilst the resolution \big(due to the long collimation section of 9.8\,m and resulting divergences of $0.0353$\,$^{\circ}$, $0.0357$\,$^{\circ}$ and $0.0365$\,$^{\circ}$\big) essentially remained unchanged. Horizontally, the beam divergence was $< 0.1$\,$^{\circ}$. 
The two-dimensional position sensitive detector \cite{Kampmann2004E845} provides a spatial resolution of $2.7$\,mm $\times$ $2.7$\,mm. 
As polarizing device a transportable NSF \big(filling pressure $p = 1.06 \pm 0.01$\,bar\big) was inserted into the neutron beam ahead of the sample and was replaced every two to three days. In this time period all the necessary measurements, i.\,e.\ the intensities of the primary beam, the transmittance of the NSFs and the reflectivity of a polarized beam from the sample, were carried out using the measurement sequence shown in figure \ref{Fg:Kreuzpaintner_1}. The need of the transmittance measurements will be explained later. The NSFs with a neutron path length of $d=5.50 \pm 0.05$\,cm were obtained from the ``Helios'' $^3$He polarization laboratory at MLZ \cite{Hutanu200714} and transported to the beamline in a dedicated magnetic device.


At the instrument the NSFs were inserted into a pre-aligned holder inside a magnetostatic cavity \big(see e.\,g.\ \cite{Petoukhov2006480,Masalovich2011012016} \big)which ensured a reproducible installation of the NSF. The magnetostatic cavity provides the necessary magnetic field of 1\,mT for the $^3$He to hold its polarization. It also contains a well characterised integrated radio frequency spin-flipper device, that, in our case, reverses the $^3$He polarization with an efficiency of $\geq 99.99\%$ \cite{Babcock2007172,Link201324}.  A magnetic guide field of approximately $1$\,mT between the magnetic box and the sample position prevented unintended loss in neutron beam polarization. Since NSFs exhibit a time dependent beam transmittance and beam polarization, the start times and durations for each of the neutron measurements were recorded. In this context, $t = 0$\,s represents the start time of the first transmittance measurement with a newly installed NSF.

The $20 \times 20$\, mm$^2$ large sample was measured in an externally applied in-plane magnetic field of 30\,mT. Because growth stage 1 of the sample only contains non-magnetic Cu on silicon, the reflectivity measurements result in identical reflectivity curves for $\uparrow$ and $\downarrow$ polarized neutrons. In contrast, growth stage 2 has a very thin, approximately 5 -- 6 monolayers, hetero-epitaxially grown magnetic Fe layer on top of the underlying Cu$_{\text{20\,nm}}$/Si$_{\text{substrate}}$. This Fe layer thickness was specifically chosen, as it represents the lowest thickness at which epitaxial Fe thin films show the reduced magnetic moment of fcc-Fe, but do have a $T_C$ above room temperature. A low magnetisation was important for benchmarking our technique, whilst room temperature magnetism was required, because no cooling mechanism was implemented into the coating setup. 

\section{Data Evaluation}
\subsection{Step 1: Normalisation of the Primary Beam Reference Spectrum}
\label{subsec:Step 1}
\begin{figure}
	\begin{center}
		\includegraphics[width=0.75\textwidth]{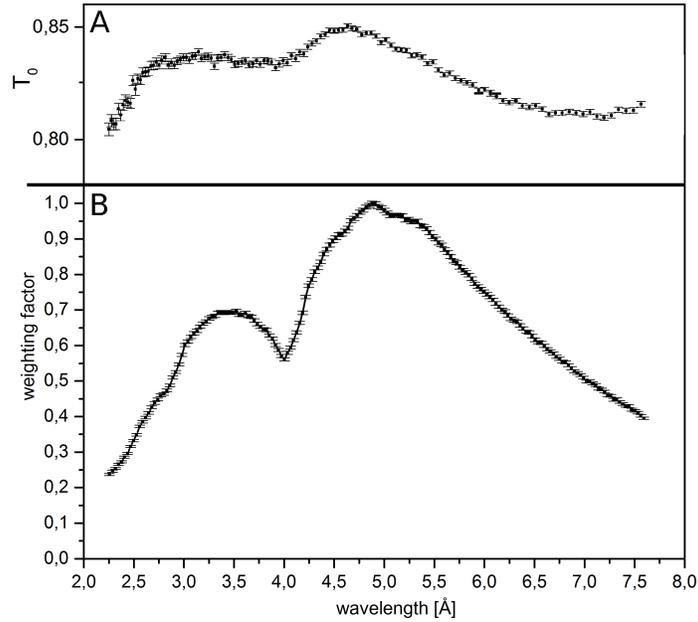}
			\caption{\textbf{A.)} The transmittance $T_0(\lambda)$ of the evacuated quartz glass $^3$He spin-filter cell. It is similar to the transmittance behaviour reported for other $^3$He spin-filter cells \cite{Chupp2007500}. \textbf{B.)} The primary beam spectrum, re-scaled to the interval of $[0, 1]$.}
			\label{Fg:Kreuzpaintner_2}
	\end{center}
\end{figure}

In order to eliminate errors that may arise due to intensity variations of the neutron source, the recorded primary beam spectra are normalised by their integral neutron intensities. 
To increase the statistics, the sum over all measured primary beam spectra was taken and re-scaled to the interval of $[0, 1]$ \big(figure \ref{Fg:Kreuzpaintner_2} B\big). Because knowledge of the absolute intensity of the primary beam spectrum is of no relevance or benefit in the proposed data treatment procedure, the fully re-scaled spectrum can be referred to and used as the primary beam spectrum $I_{\text{prim}}(\lambda)$. It will be required in nearly every consecutive step of the data treatment process, either as weighting factor or for the calculation of expectation values of neutron intensities.


\subsection{Step 2: Characterisation of the NSFs}
\label{subsec:Step 2}
\subsubsection{Background Information and \textit{a-priori} Knowledge}
\begin{figure}
	\begin{center}
		\includegraphics[width=0.75\textwidth]{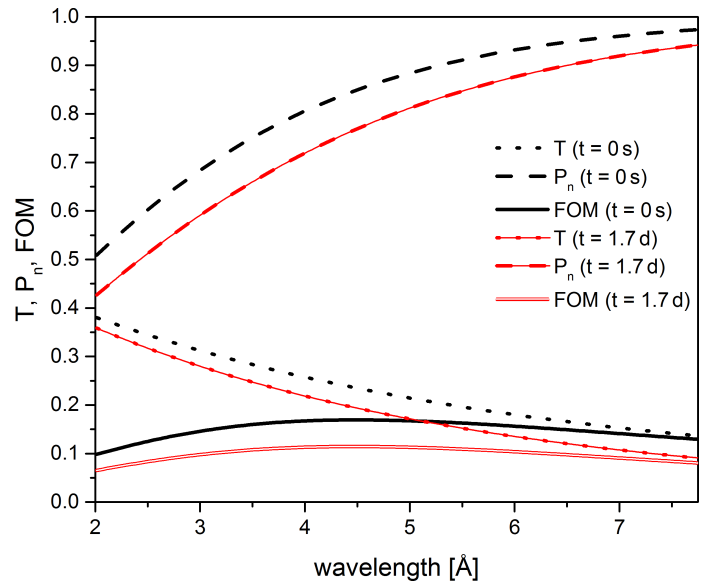}
			\caption{General polarization performance of our NSFs, immediately after insertion of a freshly prepared NSF \big($t=0$\,s, $P_{\text{He}} \sim 0.65$\big) and directly before the replacement of the NSFs \big($t=1.7$\,d, $P_{\text{He}}\sim 0.53$\big) at the end of an experimental run. Shown are the transmittance $T(\lambda,t)$, the resulting polarization $P_{\text{n}}(\lambda,t)$ of the neutron beam and the figure of merit $\text{FOM}(\lambda,t)$.}
			\label{Fg:Kreuzpaintner_2a}
	\end{center}
\end{figure}

A NSF cell is usually made of high purity quartz glass or aluminosilicate glass \big(see e.\,g.\ \cite{Chupp2007500}\big) with a high transmittance $T_0(\lambda)$, which only slightly depends on the wavelength. $T_0(\lambda)$ of the NSF used at REFSANS was measured in a separate experiment and is shown in figure \ref{Fg:Kreuzpaintner_2} A.
If filled with spin-polarized $^3$He gas with an initial polarization of approximately $65$\,\%\footnote{Since the time of the experiment, the value has been increased, such that presently 75\,\% and greater is available \cite{Link201226}}, a wavelength dependent neutron polarization of typically $> 90$\,\% \big(by demand up to 99.9\,\% but with much lower transmittance\big) can be achieved. Any $^3$He gas with polarization below $100$\,\% strongly attenuates the neutron beam due to absorption of the desirable polarized neutrons by the $^3$He atoms with the wrong polarization. Therefore, by measuring the transmittance $T(\lambda,t)$ of a NSF at different times $t$, the decay of the $^3$He polarization $P_{\text{He}}(t)$ and the corresponding degree of polarization of the neutron beam $P_{\text{n}}(\lambda,t)$ \big(see e.\,g.\ \cite{Hutanu2012012012}\big) can be recorded. The general behaviour of the transmittance $T(\lambda,t)$ and the resulting polarization $P_{\text{n}}(\lambda,t)$ of our NSFs as well as their typical polarization performance, described by the figure of merit $\text{FOM}(\lambda,t) = P_{\text{n}}^2 T$, is given in figure \ref{Fg:Kreuzpaintner_2a}.


The required mathematical relations for a detailed characterisation of a NSF are given in section \ref{subsubsec:mathematical_relationships} and exemplarily applied to our measured data in section \ref{subsubsec:determination_of_T_P}.

\subsubsection{Mathematical Relationships for Characterisation of a NSF}
\label{subsubsec:mathematical_relationships}
The transmittance of an incident unpolarized neutron beam as a function of NSF parameters is given by \big(see e.\,g.\ \cite{Hutanu2012012012,Masalovich2007791}\big)
\begin{equation}
	T(\lambda,t) = T_0(\lambda) e^{-\eta(\lambda)}\cosh\big(\eta(\lambda)\ P_{\text{He}}(t)\big)
\label{Eq:transmittance}	
\end{equation}
with the opacity
\begin{equation}
	\eta(p,d,\lambda) = 7.32 \times 10^{-2}\,p\,d\,\lambda
\label{Eq:eta}
\end{equation}	
where $t$\,[s] is the time, $\lambda$\,[\AA] is the wavelength of the neutrons, $T_0(\lambda)$ is the transmittance of the evacuated NSF cell, $d$\,[cm] is the path length, which the neutrons have to pass in the $^3$He gas and $p$\,[bar] is the filling pressure at room temperature.
$P_{\text{He}}(t)$ denotes the time dependent degree of $^3$He gas polarization from which the polarization of the neutron beam for each wavelength and point in time can be calculated as
\begin{equation}
		P_{\text{n}}(\lambda,t) = \tanh{\big(\eta(\lambda) P_{\text{He}}(t)\big)}.	
\end{equation}

\begin{figure}
	\begin{center}
		\includegraphics[width=1\textwidth]{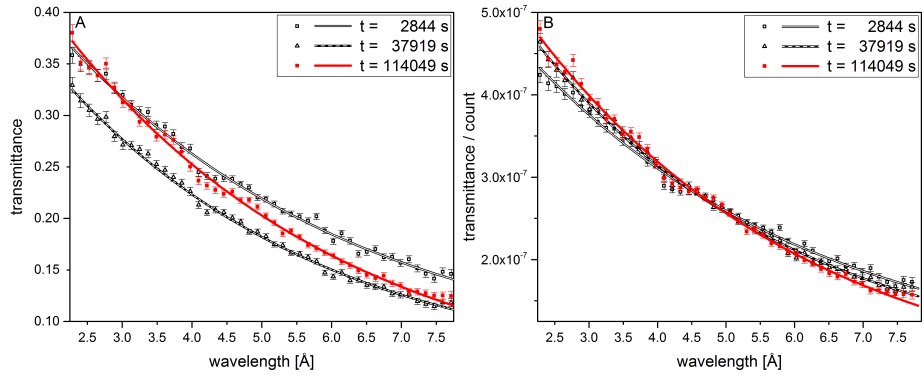}
			\caption{\textbf{A.)} Exemplary transmittance $T(\lambda,t)$ of a NSF, measured with almost fully closed collimator slits at approximately $45$\,min ($t=2844$\,s), $10.5$\,h ($t=37919$\,s) and $31.75$\,h ($t=114049$\,s) after the insertion of the NSF into the beamline (for reasons of clarity, only every fifth data point is shown. The lines are a guide to the eye). Because the remaining slit openings were of the order of the positioning accuracy of the slits themselves, the absolute transmittance $T(\lambda,t)$ is unknown. This is clearly visible by $T(\lambda,t=114049$\,s$)$ being larger than $T(\lambda,t=37919$\,s$)$, which can never be the case if a transportable NSF is used. \textbf{B.)} The same NSF transmittance measurements, but normalised by the total counts of neutrons in the measurement. This representation highlights the relevant differences in the neutron spectrum upon which our Two-Wavelength-Method is based.}
			\label{Fg:Kreuzpaintner_7c}
	\end{center}
\end{figure}

\paragraph{Two-Wavelength-Method for Extracting the $^3$He Polarization $P_{\text{He}}(t)$ From the Transmittance Measurements}
On nearly every neutron instrument the intensities of the unattenuated primary beam, required as reference for the transmittance of a NSF, saturates the neutron detector. Consequently, well thought out strategies for the measurements of absolute intensities exist. Due to limitations imposed by our sample environment (i.\,e.\ only manually controllable lateral slits), only the simplest method for protecting the neutron detector could be used, namely to perform both, the primary beam and transmittance measurements, with almost fully closed collimator slits. The remaining slit openings were then of the order of the positioning accuracy of the slits themselves. Hence, the absolute transmittance $T(\lambda,t)$ \big(equation \eqref{Eq:transmittance}\big) is scaled by an unknown slit-opening factor, if the slits needed to be moved between two consecutive transmittance measurements. Figure \ref{Fg:Kreuzpaintner_7c}\,A demonstrates the effect on the basis of three exemplary transmittance measurements, carried out after different decay times of the NSF. 


As the absolute transmittance of the NSF could not be accurately determined, any NSF characterisation which is based on the knowledge of the absolute transmittance is non-applicable. Therefore, as alternative, we developed an elegant two-wavelength-method for the extraction of $P_{\text{He}}(t)$, which eliminates the requirement of knowing the absolute transmittances, but requires knowledge of the opacity $\eta(\lambda)$, which we obtained from the $^3$He cell parameters. The method is based on the knowledge of the relative transmittances $T(\lambda,t)$ and $T(2\lambda,t)$ (figure \ref{Fg:Kreuzpaintner_7c}\,B), which are always known from within one single ToF measurement, fully independent of any potential positioning accuracies of beamline components or variations in the incoming neutron intensities:

In \cite{Masalovich2007791} a method for obtaining the $^3$He polarization $P_{\text{He}}(t)$ in a NSF of any suitable shape, which allows the transmittances $T(\eta(d))$ and $T(\eta(2d))$ for two path lengths $d$ and $2d$ through the very same NSF to be measured, was presented. Its fundamental principle is that by measuring these two transmittances \big(equation \eqref{Eq:transmittance}\big), a set of equations can be obtained which allows the $^3$He polarization $P_{\text{He}}(t)$ in the NSF to be extracted. Because equation \eqref{Eq:eta} is linear in $d$ and $\lambda$, the identity $\eta(2d,\lambda) = \eta(d,2\lambda)$ is valid. Therefore, the combination of $\eta$ and $2\eta$ can not only be fulfilled for two path-lengths $d$ and $2d$, but also for two wavelengths $\lambda$ and $2\lambda$ in the continuous spectrum.  
The extraction of $P_{\text{He}}(t)$ can be carried out analogously to the case shown in \cite{Masalovich2007791}, but by writing the equation in a different form
\begin{equation}
		Q := \frac{T(2\lambda,t)\,T_0(\lambda)}{T(\lambda,t)\,T_0(2\lambda)}\,e^{\eta(\lambda)}=\frac{1+P_{\text{n}}^2(\lambda,t)}{\sqrt{1-P_{\text{n}}^2(\lambda,t)}}.
		\label{Eq:extraction}
\end{equation}
$T(\lambda,t)$ and $T(2\lambda,t)$ are the measured transmittances of the NSF. $T_0(\lambda)$ and $T_0(2\lambda)$ are the transmittances of the empty, evacuated NSF.
Notice that the parametrization given by equation \eqref{Eq:extraction} does not require knowledge of the absolute transmittance, but only the ratio of the transmittances $T(\lambda,t)$ and $T(2\lambda,t)$. 

In this context, it should also be noted that in the case of non precisely known NSF parameters, which are required to calculate $\eta(\lambda)$, equation \eqref{Eq:extraction} offers a simple access to a measured value of $\eta(\lambda)$. For this the transmittance of the deliberately depolarised NSF (with resulting $P_n = 0$) can be measured after all other measurements have been performed. However, a reliable full depolarisation requires an additional depolarizing coil integrated into the magnetostatic cavity or the generation of a depolarizing RF pulse. For our experiments, this was not required, since $\eta(\lambda)$ can sufficiently precisely be estimated from the known NSF parameters.

Solving equation \eqref{Eq:extraction}, one obtains

\begin{equation}
P_{\text{n}}(\lambda,t)=\frac{1}{\sqrt{2}}\sqrt{Q\sqrt{8+Q^2}-Q^2-2}
		\label{Eq:extraction_of_P}
\end{equation}
from which the $^3$He polarization $P_{\text{He}}(t)$ can be calculated by

\begin{equation}
P_{\text{He}}(t)=\frac{\tanh^{-1}P_{\text{n}}(\lambda,t)}{\eta(\lambda)}.
		\label{Eq:extraction_of_P_He}
\end{equation}
As the transmittance is only measured at discrete points in time $t$, equations \eqref{Eq:extraction}, \eqref{Eq:extraction_of_P} and \eqref{Eq:extraction_of_P_He} experimentally return discrete $P_{\text{He}}(t)$ values. The full time behaviour of $P_{\text{He}}(t)$ is deductible from the \textit{a-priori} knowledge that the $^3$He polarization decays exponentially according to
\begin{equation}
	P_{\text{He}}(t) = P_0\,e^{-\Gamma t},
	\label{Tb:3He_decay_formula}
\end{equation}
where $P_0=P_{\text{He}}(t=0)$ is the initial $^3$He polarization for $t=0$\,s, $\Gamma=\frac{1}{T_1}$ [s$^{-1}$] is the relaxation rate, defined by the relaxation time $T_1$. 

\subsubsection{Experimental Determination of the Time Dependent Transmittance and Polarization of a NSF}
\label{subsubsec:determination_of_T_P}
\begin{figure}
	\begin{center}
		\includegraphics[width=0.75\textwidth]{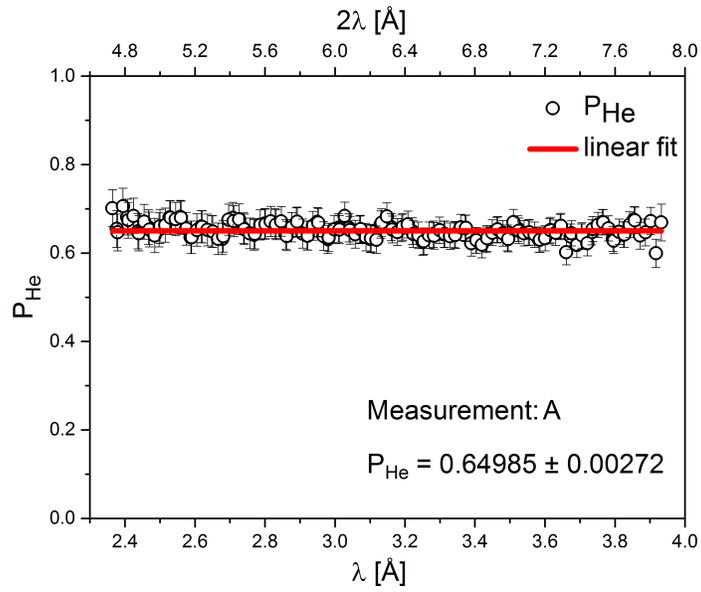}
			\caption{The $^3$He polarization $P_{\text{He}}(t)$ in measurement A, obtained using the two-wavelength-method for all data points at different pairs of wavelengths $\lambda$ and $2\lambda$, respectively. Because $P_{\text{He}}(t)$ within a single measurement must be identical, the available statistics was exploited by averaging over the ensemble of measured data points and reduce errors in the values of $P_{\text{He}}(t)$.}
			\label{Fg:Kreuzpaintner_3}
	\end{center}
\end{figure}

The transmittance $T(t,\lambda)$ of the NSF was measured alternating with reflectivity measurements. Table \ref{Tb:start_times_3He} lists the point in time at which a particular transmittance measurement was started, the integral neutron intensities detected, and the degree of $^3$He gas polarization $P_{\text{He}}(t)$, whose determination will be shown hereafter for the NSF used in measuring the reflectivities of the sample after growth stage 2 \big(figure \ref{Fg:Kreuzpaintner_3}\big). With the known time stamps, the NSF transmittance measurements can be fully related to the sample reflectivity measurements, which then can be corrected with the transmittance and polarization efficiencies delivered by the NSF.



\paragraph{Transmittance Measurements} Due to very short data acquisition times \big($600$\,s\big) of the NSF transmittance measurements, if compared to the decay time of the $^3$He polarization, it can be assumed that the measurements took place in the middle of the acquisition time span. With the mathematical relationships given in \ref{subsubsec:mathematical_relationships}, the NSF described in table \ref{Tb:start_times_3He}, can be analysed for its respective $^3$He polarization $P_{\text{He}}(t)$, from which $T(\lambda,t)$ and $P_{\text{n}}(\lambda,t)$ can be calculated.

As a first step in data treatment, the transmittance measurements A -- J \big(table \ref{Tb:start_times_3He}\big) are normalised with their corresponding integral neutron intensities. The intensities are then divided by the primary beam spectrum $I_{\text{prim}}(\lambda)$ such that the result provides the information on the relative transmittance of the NSF as function of wavelength $T(\lambda,t)$ at the corresponding points in time $t$ at which they were measured. By pairwise comparing the transmittance $T(\lambda,t)$ with the corresponding transmittance $T(2\lambda,t)$ in the measured wavelength band, the extraction of the $^3$He polarization can then be performed using the two-wavelength method \big(equations \eqref{Eq:extraction}, \eqref{Eq:extraction_of_P} and \eqref{Eq:extraction_of_P_He}\big) for each measurement A -- J. Because $P_{\text{He}}(t)$ must be identical in each measurement, no matter what pair of wavelengths was chosen, one can use the statistics and average the $^3$He polarization as illustrated in figure \ref{Fg:Kreuzpaintner_3} for measurement A. This and all other polarization values for $P_{\text{He}}(t)$ are listed in table \ref{Tb:start_times_3He}. 

To determine the parameters $P_0$ and $\Gamma$ in the final step of the characterisation of the NSF, the resulting $P_{\text{He}}(t)$ values for measurements A -- J were fitted to the exponential decay model \big(equation \eqref{Tb:3He_decay_formula}\big). With these parameters, the time behaviour of $P_{\text{He}}(t)$, and hence the transmittance $T(\lambda,t)$ and the corresponding polarization $P_{\text{n}}(\lambda,t)$ of the neutron beam, are known, which is required to evaluate the reflectivity measurements \big(section \ref{subsec:Step 3}\big).

\subsection{Step 3: Extraction and Reconstruction of Reflectivity Curves}
\label{subsec:Step 3}
\subsubsection{Background Information and \textit{a-priori} Knowledge}
The reflectivities for $0 < q_z \leq 0.13$\,\AA$^{-1}$ for each growth stage were measured for three overlapping $q_z$-ranges with different corresponding nominal angles of incidence $\omega$ and different data acquisition time intervals $[t_{\text{start}}, t_{\text{stop}}]$. The measured intensities must therefore be scaled for the process of joining all parts of the reflectivity curves. Because the measured intensities were altered by a NSF in the incoming neutron beam, this process was involved and a scaling strategy, based on the time dependent transmittance $T(\lambda,t)$ and polarization $P_{\text{n}}(\lambda,t)$, delivered by a NSF \big(section \ref{subsubsec:determination_of_T_P}\big), was developed. In this strategy, the equality $R^{\uparrow}(q_z)=R^{\downarrow}(q_z)=1$, which is valid for the $\uparrow$ and $\downarrow$ reflectivities of a sample in the regime of total reflection, is used for initial scaling \big(section \ref{subsubsec:reconstructing_neutron_intensities_total_reflection}\big). For scaling those parts of a reflectivity curve which do not contain the regime of total reflection, $R^{\uparrow}(q_z)$ and $R^{\downarrow}(q_z)$ in the overlap regime with the already reconstructed part of the reflectivity curve can then be used for determination of the subsequent scaling factors \big(section \ref{subsubsec:reconstructing_neutron_intensities_without_total_reflection}\big).
The reflectivity measurements, together with their relevant parameters for data evaluation, are summarised in table \ref{Tb:start_times_reflectivity}.


\subsubsection{Mathematical Relationships for Extraction of \texorpdfstring{$R^{\uparrow}(q_z)$}{Rup} and \texorpdfstring{$R^{\downarrow}(q_z)$}{Rdown}}
The measured neutron intensities always need to be scaled during the course of data evaluation. Therefore, the neutron intensities $I_0\vert_{t_\text{start}}^{t_\text{stop}}$ which arrive at the sample during a specific data acquisition time interval $[t_\text{start},t_\text{stop}]$ must be known. These are functions of the primary beam intensity and of the attenuation by the time dependent transmittances $T(\lambda,t)$ and $T_f(\lambda,t)$ of the NSF. The index $f$ denotes the transmittance for the spin-flipped state of the $^3$He polarization. Since the efficiency $F$ of our spin-flipper is very close to 1 \big($F \geq 0.9999$ \cite{Babcock2007172,Link201324}\big) and because only a single spin-flip was performed, it will be assumed that \big(based on an incoming unpolarized beam\big) $T_f(\lambda,t) = T(\lambda,t)$ and no further differentiation between non-spin-flipped and spin-flipped transmittance of the NSF has to be made. In the case of a high efficiency flipping, the only result of the spin-flip is the reverse of the $^3$He polarization and hence the reverse of a neutron beam polarization. In what follows, the notation $\vert_{t_\text{start}}^{t_\text{stop}}$ indicates the data acquisition time intervals with the lower \big($t_\text{start}$\big) and upper \big($t_\text{stop}$\big) limits of integration. The notation $\left[g(t)\right]_{t_\text{start}}^{t_\text{stop}}$ will be used as an abbreviation for the difference $g(t_\text{stop})-g(t_\text{start})$ of any arbitrary time dependent function $g(t)$. 
The neutron intensities reaching the sample can be calculated in units of the scaled primary beam intensities for the non-spin-flipped \big(index: +\big) and spin-flipped \big(index: -\big) state of the NSF:
\begin{eqnarray}
\label{Eq:eqnintensity+}
I_0^+\vert_{t_a}^{t_b}(\lambda) &=& I_{\text{prim}}(\lambda)\int_{t_a}^{t_b} T(\lambda,t)\,dt = -I_{\text{prim}}(\lambda)\frac{1}{\Gamma}T_0(\lambda) e^{-\eta(\lambda)}\left[\text{Chi}\,x(\lambda,t)\right]_{t_a}^{t_b}\\
I_0^-\vert_{t_c}^{t_d}(\lambda) &=& I_{\text{prim}}(\lambda)\int_{t_c}^{t_d} T(\lambda,t)\,dt = -I_{\text{prim}}(\lambda)\frac{1}{\Gamma}T_0(\lambda) e^{-\eta(\lambda)}\left[\text{Chi}\,x(\lambda,t)\right]_{t_c}^{t_d}
\label{Eq:eqnintensity-}
\end{eqnarray}
where $x(\lambda,t) = \eta(\lambda) P_0 e^{-\Gamma t}$. The quantities $I_0^+\vert_{t_a}^{t_b}(\lambda)$ and $I_0^-\vert_{t_c}^{t_d}(\lambda)$ represent calculated values used for the neutron intensities reaching the sample over a given time interval $[t_\text{a},t_\text{b}]$ and $[t_\text{c},t_\text{d}]$, respectively, in which the corresponding measurements were acquired. $\text{Chi}$ denotes the hyperbolic cosine integral which can be calculated with the use of a known series expansion \cite{Abramowitz64}. Once the relative intensities of the non-spin-flipped and spin-flipped measurements have been obtained, the following set of equations must be solved for each wavelength slice to extract $R^{\uparrow}(q_z) = R^{\uparrow}(\lambda,\alpha_i)$ and $R^{\downarrow}(q_z)=R^{\downarrow}(\lambda,\alpha_i)$:

\begin{align}
\label{Eq:eqnsystem1}
\begin{split}
		I\vert_{t_a}^{t_b}(\lambda,\alpha_i) := \frac{I^+(\lambda,\alpha_i)}{I_{\text{prim}}(\lambda)} &= \frac{1}{2} R^{\uparrow}(\lambda,\alpha_i) \int_{t_a}^{t_b}{T(\lambda,t)(1+P_{\text{n}}(\lambda,t))\,dt} \\ &+ \frac{1}{2} R^{\downarrow}(\lambda,\alpha_i) \int_{t_a}^{t_b}{T(\lambda,t)(1-P_{\text{n}}(\lambda,t))\,dt}
		\\
		I\vert_{t_c}^{t_d}(\lambda,\alpha_i) := \frac{I^-(\lambda,\alpha_i)}{I_{\text{prim}}(\lambda)} &= \frac{1}{2} R^{\uparrow}(\lambda,\alpha_i) \int_{t_c}^{t_d}{T(\lambda,t)(1-P_{\text{n}}(\lambda,t))\,dt} \\ &+ \frac{1}{2} R^{\downarrow}(\lambda,\alpha_i) \int_{t_c}^{t_d}{T(\lambda,t)(1+P_{\text{n}}(\lambda,t))\,dt}
\end{split}	
\end{align}
$I^+(\lambda,\alpha_i)$ and $I^-(\lambda,\alpha_i)$ are the scaled reflected intensities with the non-spin-flipped state and the spin-flipped state of the NSF, respectively.

Because the left hand side of the equation system \eqref{Eq:eqnsystem1} contains the measured data, only the integrals
\begin{equation}
\label{Eq:M}
M^{\pm}\vert_{t_a}^{t_b}(\lambda) := \frac{1}{2}\int_{t_a}^{t_b}{T(\lambda,t)(1 \pm P_{\text{n}}(\lambda,t))\,dt} = -\frac{T_0(\lambda)}{2\Gamma}\,e^{-\eta(\lambda)}\left[\text{Chi}\,x(\lambda,t) \pm \text{Shi}\,x(\lambda,t)\right]_{t_a}^{t_b}
\end{equation}
need to be solved using $P_{\text{He}}(t)$ as given in equation \eqref{Tb:3He_decay_formula} \big(analogously for other relevant time intervals given in table \ref{Tb:start_times_reflectivity}\big). In equation \eqref{Eq:M} $\text{Shi}$ denotes the hyperbolic sine integral of the argument \cite{Abramowitz64}.
The solution is then given by

\begin{equation}
\label{Eq:R_plus}
R^{\uparrow\downarrow}(\lambda,\alpha_i) = \frac{I\vert_{t_c}^{t_d}(\lambda,\alpha_i)M^\mp\vert_{t_a}^{t_b}(\lambda) - I\vert_{t_a}^{t_b}(\lambda,\alpha_i)M^\pm\vert_{t_c}^{t_d}(\lambda)}{M^\mp\vert_{t_a}^{t_b}(\lambda)M^\mp\vert_{t_c}^{t_d}(\lambda)-M^\pm\vert_{t_a}^{t_b}(\lambda)M^\pm\vert_{t_c}^{t_d}(\lambda)}. \\
\end{equation}

\subsubsection{Extraction of \texorpdfstring{$R^{\uparrow}(q_z)$}{Rup(qz)} and \texorpdfstring{$R^{\downarrow}(q_z)$}{Rdown(qz)} From Measurements Which Include the Regime of Total Reflection}
\label{subsubsec:reconstructing_neutron_intensities_total_reflection}
This section exemplarily demonstrates the extraction of the reflectivity curves from the two corresponding measurements I \big(+\big) and IV \big(-\big) performed after growth stage 2 with the same nominal angle of incidence $\omega = 0.25$\,$^\circ$. Because the sample is magnetic, different reflectivities for $\uparrow$ and $\downarrow$ cross sections will be obtained. The reflectivity curves of the non-magnetic growth stage 1, which do not show a splitting, were obtained analogously.

\begin{figure}
	\begin{center}
		\includegraphics[width=\textwidth]{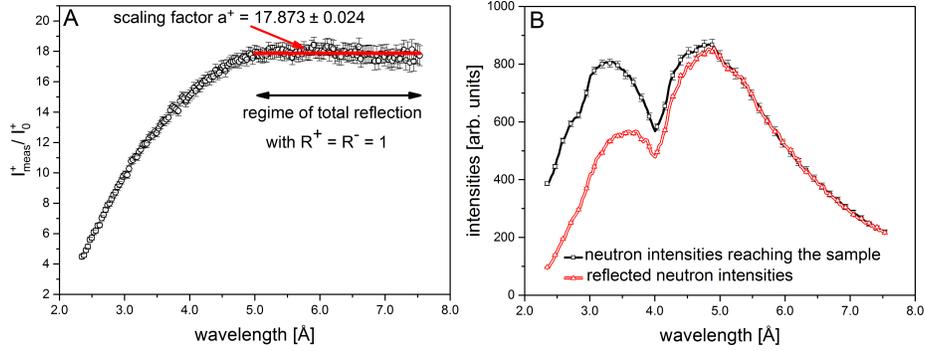}
			\caption{A.) Determination of the scaling factor $a^+$, required to scale the intensities of reflectivity measurement I to an intensity level, which is defined by the primary neutron intensities reaching the sample $I_0^+\vert_{t_0}^{t_1}(\lambda)$. B.) The overlay of the scaled intensities and the incoming intensities (for reasons of clarity, only every fifth data point is shown): only the scaling of the measured intensities into relative intensities $I^+\vert_{t_0}^{t_1}(\lambda)=\frac{1}{a^+}I_{\text{meas}}^+\vert_{t_0}^{t_1}(\lambda)$ \big(in combination with the analogously rescaled $I^-\vert_{t_6}^{t_7}(\lambda)=\frac{1}{a^-}I_{\text{meas}}^-\vert_{t_6}^{t_7}(\lambda)$\big) allows the $\uparrow$ and $\downarrow$ neutron reflectivities $R^{\uparrow}(q_z)$ and $R^{\downarrow}(q_z)$ to be extracted.
}
			\label{Fg:Kreuzpaintner_4}
	\end{center}
\end{figure}

\paragraph{Scaling of the Neutron Intensities}
To rule out any unintended variations in the measured neutron intensities $I\vert_{t_0}^{t_1}(\lambda)_{\text{meas}}$ and $I\vert_{t_6}^{t_7}(\lambda)_{\text{meas}}$ \big(table \ref{Tb:start_times_reflectivity} \big), the first step in data treatment lies with the scaling to relative neutron intensity values. 
The starting point in this process is the calculation of the neutron intensities $I_0\vert_{t_0}^{t_1}(\lambda)$ and $I_0\vert_{t_6}^{t_7}(\lambda)$, which arrived at the sample for the corresponding data acquisition time intervals $[t_0,t_1]$ and $[t_6,t_7]$ \big(equations \eqref{Eq:eqnintensity+} and \eqref{Eq:eqnintensity-}, $P_0$ and $\Gamma$ are given in table \ref{Tb:start_times_3He}\big). 
The required scaling factor $a^+$ for measurement I \big($a^-$ for measurement IV\big) can be obtained from the intensities in the wavelength range, which constitutes the regime of total reflection: In a first step, $I\vert_{t_0}^{t_1}(\lambda)_{\text{meas}}$ is divided by $I_0\vert_{t_0}^{t_1}(\lambda)$ \big(analogously $I\vert_{t_6}^{t_7}(\lambda)_{\text{meas}}$ by $I_0\vert_{t_6}^{t_7}(\lambda)$\big). The resulting graph is shown in figure \ref{Fg:Kreuzpaintner_4}\,A. It gives the unscaled total reflectivity of the sample as a function of wavelength. Then, using the common scaling condition $R^{\uparrow}(q_z)=R^{\downarrow}(q_z)=1$, in the wavelength range in which total reflection occurs \big($\lambda > 5$\,\AA\big) and by averaging over all discrete data points in that range, a scaling factor $a^+=17.873 \pm 0.024$ is obtained \big(and $a^-=17.880 \pm 0.026$. The slightly differing scaling factors $a^+$ and $a^-$ may be obtained due to tolerances in the slit openings and resulting sample illumination\big). With the known scaling factors the reflected neutron intensities can be scaled to relative intensity values $I\vert_{t_0}^{t_1}(\lambda)=\frac{1}{a^+}I\vert_{t_0}^{t_1}(\lambda)_{\text{meas}}$ and $I\vert_{t_6}^{t_7}(\lambda)=\frac{1}{a^-}I\vert_{t_6}^{t_7}(\lambda)_{\text{meas}}$ \big(figure \ref{Fg:Kreuzpaintner_4}\,B\big).


This allows the extraction of the $\uparrow$ and $\downarrow$ neutron reflectivities $R^{\uparrow}(q_z)$ and $R^{\downarrow}(q_z)$ (equation \eqref{Eq:R_plus}) to be carried out.

\subsubsection{Extraction of \texorpdfstring{$R^{\uparrow}(q_z)$}{Rup(qz)} and \texorpdfstring{$R^{\downarrow}(q_z)$}{Rdown(qz)} From Measurements Which Do Not Include the Regime of Total Reflection}
\label{subsubsec:reconstructing_neutron_intensities_without_total_reflection}
This section shows how the extraction of the reflectivity curves from the two corresponding measurements II \big(+\big) and V \big(-\big) of growth stage 2, taken at identical nominal angles of incidence $\omega=0.6$\,$^{\circ}$, is performed. Because only the scaling procedure differs from the one presented in section \ref{subsubsec:reconstructing_neutron_intensities_total_reflection}, whilst the remaining process of extracting the reflectivities is identical, only the former will be described.

\paragraph{Scaling of the Neutron Intensities}
As for all previously described measurements, $I\vert_{t_2}^{t_3}(\lambda)_{\text{meas}}$ and $I\vert_{t_8}^{t_9}(\lambda)_{\text{meas}}$ must undergo a scaling process, this time with the primary goal of comparing the intensities to those ones taken at a nominal angle of incidence $\omega = 0.25$\,$^\circ$. The required scaling factor $b^+$ for measurement II \big(and $b^-$ for measurement V\big) can be calculated recursively from the reflectivities $R^{\uparrow}(q_z)_{\text{prev}}$ and $R^{\downarrow}(q_z)_{\text{prev}}$ in the overlap regime with the previous part of the reflectivity curve. The index $\text{prev}$ is used when values of the already reconstructed part of the reflectivity curve are explicitly referred to. If no index is given, the values generally correspond to the exemplarily treated measurements II and V.

\begin{figure}
	\begin{center}
		\includegraphics[width=0.75\textwidth]{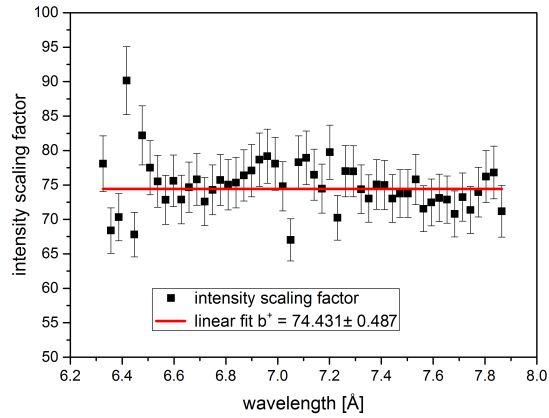}
			\caption{In order to perform the extraction of reflectivities in a way that allows the second part of the reflectivity curve \big(measurement II\big) to be joined to the first part \big(measurement I\big), the measured intensities must be scaled to the level of expected intensities by the scaling factor $b^+$.}
			\label{Fg:Kreuzpaintner_6}
	\end{center}
\end{figure}

The starting point in the scaling process is to calculate the integrals $M^{\pm}\vert_{t_2}^{t_3}(\lambda)$ and $M^{\pm}\vert_{t_8}^{t_9}(\lambda)$ and to obtain the reflectivities $R^{\uparrow}(q_z)$ and $R^{\downarrow}(q_z)$ of measurements II and V by interpolation of the reflectivities $R^{\uparrow}(q_z)_{\text{prev}}$ and $R^{\downarrow}(q_z)_{\text{prev}}$ in the overlap regime. 
As for each $q_z$ the corresponding pairs ($\lambda, \alpha_i$) and the corresponding $M^{\pm}\vert_{t_2}^{t_3}(\lambda)$ and $M^{\pm}\vert_{t_8}^{t_9}(\lambda)$ are known, expectation values $I\vert_{t_2}^{t_3}(\lambda)_{\text{expect}}$ and $I\vert_{t_8}^{t_9}(\lambda)_{\text{expect}}$ can be calculated using equations \eqref{Eq:eqnsystem1}. The required scaling factor $b^+$ for measurement II \big($b^-$ for measurement V\big) can be obtained by dividing the measured intensities $I\vert_{t_2}^{t_3}(\lambda)_{\text{meas}}$ by the corresponding calculated expected intensities $I\vert_{t_2}^{t_3}(\lambda)_{\text{expect}}$ \big(analogously for $I\vert_{t_8}^{t_9}(\lambda)_{\text{meas}}$ and $I\vert_{t_8}^{t_9}(\lambda)_{\text{expect}}$\big). As small variations of the scaling factor $b^+$ with $\lambda$ occur \big(likewise for $b^-$\big), the scaling factor is averaged over the scaling factors for all wavelengths in the overlap regime \big(figure \ref{Fg:Kreuzpaintner_6}\big).


A scaling factor of $b^+=74.431 \pm 0.487$ \big($b^-=75.185 \pm 0.558$\big) is obtained. With the known scaling factors the reflected neutron intensities can be scaled to relative intensity values $I\vert_{t_2}^{t_3}(\lambda)=\frac{1}{b^+}I\vert_{t_2}^{t_3}(\lambda)_{\text{meas}}$ and $I\vert_{t_8}^{t_9}(\lambda)=\frac{1}{b^-}I\vert_{t_8}^{t_9}(\lambda)_{\text{meas}}$ and the extraction of the $\uparrow$ and $\downarrow$ neutron reflectivities $R^{\uparrow}(q_z)$ and $R^{\downarrow}(q_z)$ can be carried out analogously to the process described in section \ref{subsubsec:reconstructing_neutron_intensities_total_reflection}. The resulting reflectivities extend the previous parts of the reflectivity curves to larger $q_z$.

The process of scaling and appending can successively be repeated for all remaining reflectivity measurements \big(III and VI\big) until the reflectivity curves $R^{\uparrow}(q_z)$ and $R^{\downarrow}(q_z)$ over the full measured range have been re-constructed. A re-binning of the data points in $q_z$ with equidistant $\Delta q_z$ \big(in the presented case $\Delta q_z=0.002$\,\AA$^{-1}$\big), completes the process of data-treatment and makes the polarized reflectivity curves available for the scientific analysis.

\section{Reliability Check and Concluding Remarks}
\begin{figure}
	\begin{center}
		\includegraphics[width=0.75\textwidth]{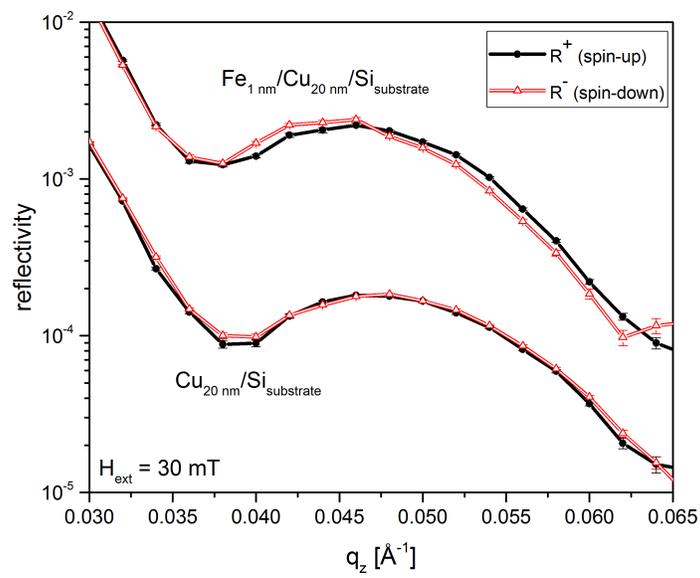}
			\caption{Detailed excerpt of the reconstructed reflectivity curves as measured with polarized neutrons for the sample growth stages Fe$_{\text{1\,nm}}$/Cu$_{\text{20\,nm}}$/Si$_{\text{substrate}}$ \big(top curves\big) and Cu$_{\text{20\,nm}}$/Si$_{\text{substrate}}$ \big(lower curves, shifted by a factor of 10 for clarity\big). The $R^{\uparrow}(q_z)$ and $R^{\downarrow}(q_z)$ data points in the shown $q_z$ range are connected by lines as guide to the eyes.}
			\label{Fg:Kreuzpaintner_7a}
	\end{center}
\end{figure}

\begin{figure}
	\begin{center}
		\includegraphics[width=0.75\textwidth]{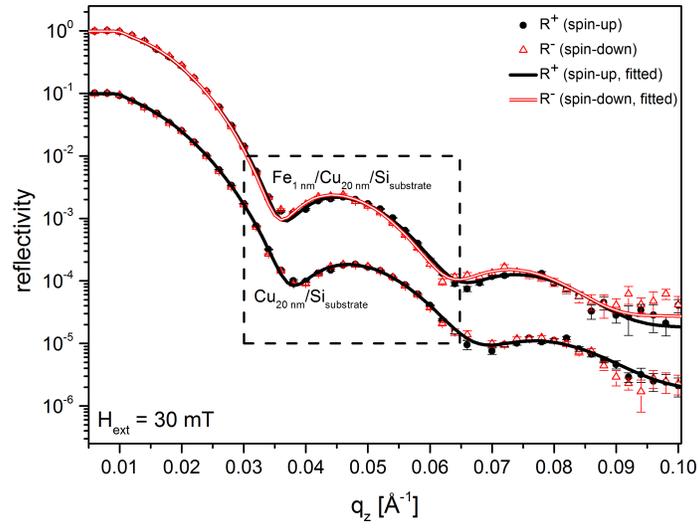}
			\caption{Reflectivity curves in the full $q_z$-range measured for the sample growth stages Fe$_{\text{1\,nm}}$/Cu$_{\text{20\,nm}}$/Si$_{\text{substrate}}$ \big(top curves\big) and Cu$_{\text{20\,nm}}$/Si$_{\text{substrate}}$ \big(lower curve, shifted by a factor of 10 for clarity\big). The model for theoretically fitted reflectivity curves yield a magnetisation of $0.6 \mu_{\text{Bohr}}$ per unit cell for the Fe layer. The dashed frame indicates the $q_z$-range, displayed in figure \ref{Fg:Kreuzpaintner_7a}.}
			\label{Fg:Kreuzpaintner_7b}
	\end{center}
\end{figure}

The data evaluation process described in the previous paragraphs was also applied to the measurements of the non-magnetic sample after growth stage 1, which does not show a splitting in the spin-channels after the identical data treatment. The resulting reflectivity curves for both growth stages are shown in figure \ref{Fg:Kreuzpaintner_7a} in a limited $q_z$-range for better visibility. The full reflectivity curves together with fitted theoretical models, which assume a magnetisation of $0.6 \mu_{\text{Bohr}}$ per atom for the Fe layer after growth stage 2 are given in figure \ref{Fg:Kreuzpaintner_7b}. The model fitting to the measured data was performed using SimulReflec \cite{Simulreflec}. A good agreement between the measured data and simulated curves confirms the reliability of the presented measurement and data extraction procedures.


In summary, we have shown that a ToF neutron reflectometer can easily be upgraded to provide a polarized neutron beam by simply placing a NSF in a magnetostatic cavity with an integrated spin-flipper system directly in front of the sample position. In this context, our aim was to draw an attention to the fact that pure mathematics alone allows one to perform a data evaluation when using a NSF with as little as one spin-flip process. The presented experimental procedure and the measurement sequence proved very suitable for the exemplary extraction of the $\uparrow$ and $\downarrow$ reflectivities $R^{\uparrow}(q_z)$ and $R^{\downarrow}(q_z)$. The presented evaluation strategy for the data, together with the mathematical treatment, constitute the basis for the simplicity in the experimental procedure. Possible simplifications of the mathematical approach are currently being investigated. The processing of the data will also be developed further for automation and implementation within a computer program, which will be presented elsewhere, once most of the necessary steps for data treatment can be performed fast and without user interaction. The presented approach was developed with the aim of aiding the neutron user community in the use of NSFs on todays ToF neutron reflectometers, where due to the available neutron flux, the data acquisition times required for sample measurements cannot be considered as being short when compared with the decay time of a $^3$He polarization. The two-wavelength method presented in section \ref{subsec:Step 2} constitutes a simple fail-safe and universally valid characterisation method for NSFs with a broad range of applications on all types of neutron scattering setups \big(e.\,g.\ small angle scattering setups (SANS), spectrometers or diffractometers\big). Most importantly, it can be applied even in the case when the absolute transmittance of the NSF cannot be determined. We also expect it to be of benefit on future high-flux neutron sources to characterise a NSF in single shot measurements, which are then immediately followed by short sample measurements, where the decay in $^3$He polarization can be neglected. It should also be noted, that the two-wavelength method is not limited to neutron scattering setups using ToF. It can even give a practical application to the usually unwanted \big(and typically filtered\big) $\frac{1}{2}\lambda$ contamination in monochromatic neutron instruments to fulfill the two-wavelength condition for the characterisation of a NSF.

\section*{Acknowledgements}
The funding of this project by Deutsche Forschungsgemeinschaft \big(DFG\big) within the Transregional Collaborative Research Center TRR 80 ``From electronic correlations to functionality'' is gratefully acknowledged.
This work is based upon experiments performed at the REFSANS instrument operated by HZG at the Heinz Maier-Leibnitz Zentrum (MLZ), Garching, Germany.

\section*{References}
\bibliography{Kreuzpaintner}
\newpage



\begin{table}
			\caption{Protocol of the parameters of the NSF used for calculating the reflectivities of the sample after growth stage 2. Listed are the scan numbers of the transmittance measurements [\#] and the corresponding point in time $t$ [s] at which the transmittance measurement was started, the data acquisition times $\Delta t$ [s], the integral neutron intensities $I$ [cnts] and the calculated $^3$He gas polarization $P_{\text{He}}(t)$ obtained from equation \eqref{Eq:extraction_of_P_He}. The bottom line shows the obtained fit parameters $P_0$ and $\Gamma$ according to equation \eqref{Tb:3He_decay_formula}. The uncertainty is given by the standard deviation $\sigma$ and is the statistical error.}
			\
			\begin{center}
			\begin{tabular}{c|c|c|c|c}
measure-  & $t$  			& $\Delta t$ 	& $I$    			& $P_{\text{He}}(t)$ 		\\
ment [\#] & [s]	      &  [s]	  	  & [cnts] 			&                    		\\
\hline
\hline
A 				&      0 & 600 				& 827288 & 0.64985 $\pm$ 0.00272 \\
B 				&   2536 & 600 				& 924229 & 0.66181 $\pm$ 0.00302 \\
C 				&   7275 & 600 				& 880352 & 0.64707 $\pm$ 0.00238 \\
D 				&  26008 & 600 				& 785218 & 0.62261 $\pm$ 0.00290 \\
E 				&  98889 & 600 				& 721265 & 0.56550 $\pm$ 0.00296 \\
F 				&  99562 & 600 				& 718972 & 0.56556 $\pm$ 0.00272 \\
G 				& 102092 & 600 				& 801808 & 0.57662 $\pm$ 0.00330 \\
H 				& 106829 & 600 				& 767220 & 0.56619 $\pm$ 0.00296 \\
I 				& 125564 & 600 				& 692649 & 0.54549 $\pm$ 0.00272 \\
J 				& 150180 & 600 				& 676260 & 0.52809 $\pm$ 0.00324 \\
			\hline
			\end{tabular}
			\begin{tabular}{c c}
			&\\
			$P_0$ = 0.6540 $\pm$ 0.0062 	& $\Gamma = (1.406 \pm 0.118) 10^{-6}$\,s$^{-1}$ \\
			\hline
			\end{tabular}
			\label{Tb:start_times_3He}
			\end{center}
\end{table}

\begin{table}
			\caption{The nominal angles of incidence $\omega$ and the start and stop times of the reflectivity measurements in seconds.}
			\
			\begin{center}
			\begin{tabular}{||c||c|c|c||c|c|c||}
			        \hline
												& \multicolumn{3}{c||}{non-spin-flipped \big(+\big)}       & \multicolumn{3}{c||}{spin-flipped \big(-\big)}        			\\
				\hline
				$\omega$    		& measure-  &   start   &   stop  	& measure-   & start     & stop   							\\
				$[^{\circ}]$		& ment [\#] &   [s]	   	&   [s]			& ment [\#]  & [s]	     & [s]    							\\
				\hline
				0.25       			&  I      	&  $t_0 =  3204$ &  $t_1 = 7204$ & IV   & $t_6 = 102760$ 		& $t_7 = 106760$         \\  
				0.60       			&  II     	&  $t_2 =  7934$ &  $t_3 = 25934$ & V   & $t_8 = 107491$ 		& $t_9 = 125491$         \\
				1.40       			&  III      &  $t_4 = 26757$ &  $t_5 = 98757$ & VI  & $t_{10} = 126315$ & $t_{11} = 149985$      \\ 
				\hline
			\end{tabular}
			\end{center}
			\label{Tb:start_times_reflectivity}
\end{table}

\end{document}